\documentstyle[preprint,aps]{revtex}
\begin{document}
\draft
\title{
Ballistic transport:
A view from the quantum theory of motion
}
\author{Hua Wu and D. W. L. Sprung}
\address{
  Department of Physics and Astronomy, McMaster University\\
  Hamilton, Ontario L8S 4M1 Canada
}
\date{Submitted to Phys. Lett. A on April 1994}
\maketitle
\begin{abstract}
Ballistic transport of electrons through a quantum wire with a
constriction is studied in terms of Bohm's interpretation of
quantum mechanics, in which the concept of a particle orbit is
permitted. The classical bouncing ball trajectories, which
justify the name ``ballistic transport'', are established in the
large wave number limit. The formation and the vital role of
quantum vortices is investigated.
\end{abstract}

The ballistic transport of electrons in quantum devices based on
the two-dimensional electron gas is a subject of intense study. The
name ``ballistic transport'' is borrowed from classical mechanics
where a point-like particle moves freely in a region except for
elastic collisions with obstacles or the confining walls.
Obviously, the classical orbits are straight line segments
connected through a reflection law. Quantum devices based on
ballistic transport can be approximately thought of as an
electron waveguide, in which the electrons behave as
non-interacting particles of effective mass $m^*$ in a constant
potential with hard wall confinement.  While the picture of
bouncing balls is a vivid one,  quantum mechanically we need a
precise justification for this picture. For this purpose, we study
here a very simple model, that  of a quantum wire with a
constriction, to see how classical orbits emerge from the
quantum wavefunction.

We consider a straight quantum wire of width $d$. At the origin
$x=0$ we place a thin but opaque obstacle attached to one side
wall so that the channel opening is narrowed to $a$. Electrons of
energy $E$ are injected from the left in transverse mode $n_i$.
In order to develop our argument, we solve this problem by the
mode matching method.

Denote the wavefunction to the left and right of the narrow
constriction as $\Psi_1$ and $\Psi_2$ respectively. These can be
expanded in plane waves as
 \begin{equation}
\Psi_\Gamma = \sum_n (C_{\Gamma n} e^{i\alpha_n x} +
\overline{C}_{\Gamma n} e^{-i\alpha_n x} )\sin (k_n y)
\label{eq:1}
\end{equation}
where $\Gamma=1,2$, $k_n=n\pi/d$, and
 \begin{equation}
\alpha_n=\sqrt{E-k_n^2}
\end{equation}
with appropriate units such that $\hbar^2/(2m^*)=1$. We also define
symbols $C_{\Gamma n}^\pm = C_{\Gamma n} \pm \bar{C}_{\Gamma n}$.
At $x=0$, the wavefunction in the opening is expanded as
 \begin{equation}
\mbox{sn}(y)\equiv \Psi|_{x=0}=\left\{ \begin{array}{ll}
		    \sum_m D_m\sin\frac{m\pi y}{a}\;&0<y<a\\
                          \\
	            0&a<y<d
		    \end{array}
	     \right. \label{eq:sn}
\end{equation}
Continuity of the wavefunction at $x=0$ requires that
 \begin{equation}
C_{1n}^+ = C_{2n}^+  = \sum_m I_{nm} D_m\label{eq:v}
\end{equation}
where
 \begin{eqnarray}
I_{nm}&=&\frac{2}{d}\int_0^a \sin \frac{n\pi y}{d}\sin\frac{m\pi y}{a}
dy \nonumber\\
&=&\frac{2}{d}\left[ \frac{\sin(\frac{n\pi}{d}-\frac{m\pi}{a})a}
                          {\frac{n\pi}{d}-\frac{m\pi}{a}       }
   -                 \frac{\sin(\frac{n\pi}{d}+\frac{m\pi}{a})a}
                          {\frac{n\pi}{d}+\frac{m\pi}{a}       }
              \right] \label{eq:I}
\end{eqnarray}
Similarly, continuity of the wavefunction derivative at $x=0$
results in
 \begin{equation}
\sum_n C_{1n}^- \alpha_n I_{nm} =
\sum_n C_{2n}^- \alpha_n I_{nm} \label{eq:d}
\end{equation}
In matrix form, Eqs. (\ref{eq:v}) and (\ref{eq:d}) read:
 \begin{eqnarray}
&&C_1^+ = C_2^+  = ID\\
&&I^t\alpha C_1^- =  I^t\alpha C_2^-
\label{eq:dm}
\end{eqnarray}
Notice that the above equation does not imply $C_1^- = C_2^-$, since
$I$ is not invertible.

Physically, the scattering boundary conditions require that
$\overline{C}_2=0$ so that no backward wave is traveling in
region 2; therefore
 \begin{equation}
C_2=ID\;\;\mbox{and}\;\;\overline{C}_1=ID-C_1. \label{eq:C}
\end{equation}
Inserting this into Eq. (\ref{eq:dm}), we obtain
 \begin{equation}
D = (I^t\alpha I)^{-1}I^t\alpha C_1 \label{eq:D}
\end{equation}
Thus the wavefunction is determined by the injection mode expansion
coefficients $C_{1n}=\delta_{nn_i}$.

Calculating the transmission coefficient is not the main purpose
of this paper. Rather, we wish to examine how the electron passes
through the channel, and to compare this with the corresponding
classical motion. To this end, we adopt the formalism of
Bohm's interpretation of quantum mechanics\cite{Bohm52,Bohm87},
in which a system is described not only by its wavefunction, but
also particle orbits. In this picture, a particle at position
${\bf r}$ moves with velocity ${\bf v}={\bf j}/|\Psi|^2$, where
${\bf j}$ is the probability current density. Since Bohm's
picture is most vividly applied to scattering states, Holland in
his recent book called it ``the quantum theory of
motion''\cite{Holland}.  We are dealing with a time independent
wave equation, so the process of finding the particle
trajectories is separated from solving for the wavefunction. All
we have to do is to find the stream lines of the velocity field
deduced from the wavefunction; they are curves tangent to ${\bf
j}$.

According to our recent discussion of quantum probability
flow\cite{Wu93,Wu94}, the stream lines can be classified in three
categories: 1) those that close on themselves; 2) those which
participate in global transmission, (which start and end at
infinity; they could be regarded as special stream lines of type
1 which close at infinity;) 3) those which connect saddle points
or end on the boundary; these lines separate different pieces of
the flow. In order to study particle transport we concentrate on
type 2 stream lines as shown in Fig. 1. We drew these lines by
starting from points along the channel opening at $x=0$, and
tracing the trajectories forward and backward. The density of the
starting points was taken to be proportional to the probability
density in the opening. In order to exhibit the relation to the
classical orbits, we deliberately chose a high electron energy.
The trajectory in the transmitted region, as expected, clearly
resembles that of a classical ball bouncing along between
hard walls, or light rays between parallel mirrors. We can
understand this better by developing an approximate analytical
solution.

Consider a situation where the transverse and longitudinal wave
numbers are all large. The computed transmission coefficient is very
close to the classical prediction, $a/d$. A large transverse wave
number implies that $I_{nm}$ in Eq. (\ref{eq:I}) will have appreciable
non-zero values only when $n\pi/d\approx m\pi/a$. From this condition
and Eq. (\ref{eq:D}), we see that $D_m$ is concentrated in a narrow
interval centering about $m=n_i(a/d)$. Also from Eq.  (\ref{eq:C}),
$C_{2n}$ and $\overline{C}_{1n}$ are concentrated about $n=n_i$. These
estimates are well verified by exact numerical calculations. Denoting
$k_x=\alpha_{n_i}$, $k_y=k_{n_i}$, $\alpha_n$ can be expanded about
$k_x$:
 \begin{equation}
\alpha_n-k_x = -  (k_n -k_y )\tan\theta
\end{equation}
where
 \begin{equation}
\tan\theta=k_y/k_x \quad .
\end{equation}
Here we have used the condition of large longitudinal wave
number. Under this approximation,
 \begin{eqnarray}
\Psi_2&\approx&\exp[i(k_x+ k_y\tan\theta)x]\times\nonumber\\
&&\sum_n C_{2n}\exp(-i k_n x\tan\theta) \sin (k_n y) \nonumber\\
&=&\exp[i(\alpha_{n_i}+ k_{n_i}\tan\theta)x]\times\nonumber\\
&&\sum_n C_{2n}\left[\frac{\cos k_n(y-x\tan\theta)-\cos
k_n(y+x\tan\theta)}{2i} \right. \nonumber\\
&&\left.
+ \frac{\sin k_n(y-x\tan\theta)+\sin
k_n(y+x\tan\theta)}{2}\right] \quad .
\end{eqnarray}
Expressing $\cos k_n (y\pm x\tan\theta)$ as $(1/k_n) \partial[ \sin k_n
(y\pm x\tan\theta)]/ \partial y$, and approximating $1/k_n=1/k_y$ in
the summation, we obtain
 \begin{eqnarray}
\Psi_2&=&\exp[i(k_x+ k_y\tan\theta)x]\times\nonumber\\
&&\left\{ \frac{1}{2ik_y}\frac{\partial}{\partial y}\left[
\mbox{sn}(y-x\tan\theta) - \mbox{sn}(y+x\tan\theta)\right]\right.\nonumber\\
&+&\left.\frac{1}{2}\left[\mbox{sn}(y-x\tan\theta) +
\mbox{sn}(y+x\tan\theta)\right]
\right\}  \label{eq:sn2}
\end{eqnarray}
where the function sn($y$) is defined by Eq. (\ref{eq:sn}),
together with its odd symmetry extension, and its periodic
extension with period $2d$. In other words,
 \begin{equation}
\mbox{sn}(y)=\left\{ \begin{array}{ll}
 \sum_m D_m \sin\frac{m\pi\bar{y}}{a}\;&|\bar{y}|<a \\
  \\
  0&|\bar{y}|>a
		    \end{array} \right.
\end{equation}
where $\bar{y}$ is $y$ minus an integer multiple of $2d$
such that the result falls in $(-d,d]$. Now return to Eq.
(\ref{eq:sn2}) and consider the derivative of the function
$\mbox{sn}(y)$. Its non-zero part is $\sum_m D_m (m\pi/a)
\cos[m\pi\bar{y}/a]$. Now since $m$ is in a narrow
region of $n_ia/d$, the factor $m\pi/a$ may be
approximated by $k_y$ (this kind of approximation would not
be permitted when the factor is inside a rapid changing function
such as $\sin$, $\cos$, and exponential). This leads to
 \begin{eqnarray}
\Psi_2&\approx&\exp[i(k_x+ k_y\tan\theta)x]\times\nonumber\\
&&\frac{1}{2i}\left[
\mbox{cs}(y-x\tan\theta) - \mbox{cs}(y+x\tan\theta)\right.\nonumber\\
&+&\left.i\,\mbox{sn}(y-x\tan\theta) +
i\,\mbox{sn}(y+x\tan\theta)\right]\nonumber\\
&=&\exp[i(k_x+ k_y\tan\theta)x]\times\nonumber\\
&&\frac{\mbox{ep}[i(y-x\tan\theta)] - \mbox{ep}[-i(y+x\tan\theta)]}{2i}
  \label{eq:sn3}
\end{eqnarray}
where
 \begin{equation}
\mbox{cs}(y)=\left\{\begin{array}{ll}
 \sum_m D_m \cos\frac{m\pi\bar{y}}{a}\;&|\bar{y}|<a \\
  \\
  0&|\bar{y}|>a
		    \end{array} \right.
\end{equation}
and
 \begin{equation}
\mbox{ep}(iy)= \mbox{cs}(y) + i \mbox{sn}(y) \, .
\end{equation}
Aside from a phase factor, Eq.(\ref{eq:sn3}) expresses $\Psi_2$
in terms of an exponential-like function $\mbox{ep}(iy)$. When
$x=0$, it clearly reproduces Eq. (\ref{eq:sn}) since
$\mbox{sn}(y)=[\mbox{ep}(iy) - \mbox{ep}(-iy)]/(2i)$. As $x$
increases, $\mbox{ep}[i(y-x\tan\theta)]$ represents a function
$\mbox{ep}[iy]$ shifted by $x\tan\theta$ upwards, and
$\mbox{ep}[-i(y+x\tan\theta)]$ represents a function $\mbox{ep}[-
iy]$ shifted the same amount downwards. In other words, the two
functions $\mbox{ep}(\pm iy)$ each move at an angle $\pm\theta$
to the channel walls as $x$ increases. The motion of the
functions leaves stripes of non-zero parts of the wavefunction.
Fig. 2 illustrates the formation of the bouncing ball pattern in
between the two thick horizontal lines which represent the wire
walls. At the left edge of the figure, vertical line segments
locate the non-zero parts of the function $\mbox{ep}(\pm iy)$.
The motion of these vertical lines traces out the upward and
downward inclined stripes. What has physical meaning are those
portions where the stripes lie inside the wire. Regions marked by
a right upward arrow (U-regions) are those due to $\mbox{ep}(iy)$,
and those marked with a right downward arrow (D-regions) are due to
$\mbox{ep}(-iy)$. The triangular regions marked with horizontal
arrows (H-regions) are where the two contributions superpose.

Not only can we show that the non-zero part of the wavefunction
produces the pattern shown in Fig. 2, we can also show that the
trajectories indeed follow the directions shown. To see this we
calculate the particle velocity
 \begin{equation}
{\bf v}=(\hbar/m^*)\mbox{Im}(\Psi^*\nabla \, .
\Psi)/|\Psi|^2 \end{equation} To compute the gradient of
$\Psi_2$, first note that
 \begin{equation}
\frac{\partial \mbox{ep}(iy)}{\partial y}=ik_y\mbox{ep}(iy)
\end{equation}
in the approximation $m\pi/a\approx k_y$ that we have used
before. Thus $\mbox{ep}(iy)$ behaves much like $\exp(ik_y y)$ in the
non-vanishing region. Consider a U-region, where
$\Psi_2=\exp[i(k_x+k_y\tan\theta)x] \mbox{ep}[i(y- x\tan\theta)]/(2i)$,
 \begin{equation}
\frac{\partial \Psi_2}
{\partial x}=ik_x\Psi_2,\quad \frac{\partial
\Psi_2}{\partial y}=ik_y \Psi_2 \end{equation} which lead to
 \begin{equation}
v_x=(\hbar/m^*)k_x,\quad
v_y=(\hbar/m^*)k_y  \, ;
\end{equation}
therefore
 \begin{equation}
v_y/v_x=\tan\theta \, .
\end{equation}
That is to say, the particle moves to the right and upward in a
direction determined by the decomposition of the total wave
number in the $x$ and $y$ directions. Similarly the particle in a
D-region moves in the direction $v_y/v_x=-\tan\theta$.

In an H-region, it is easily seen that $v_x$ takes the same value as in
the previous two cases. For $v_y$, one finds
 \begin{eqnarray}
&&\Psi_2^*\frac{\partial \Psi_2}{\partial y}=\nonumber\\ &&ik_y/4
\left\{\mbox{ep}^*[i(y-x\tan\theta)] -
\mbox{ep}^*[-i(y+x\tan\theta)] \right\}\times 	\nonumber\\
&&\phantom{ik_{n_i}/4}\left\{\mbox{ep}[i(y-x\tan\theta)] +
\mbox{ep}[-i(y+x\tan\theta)] \right\}
\end{eqnarray}
which yields
 \begin{eqnarray}
&&\mbox{Im}\left(\Psi_2^*\frac{\partial \Psi_2}{\partial y}\right)
=\nonumber\\
&&k_y/4\left\{|\mbox{ep}[i(y-x\tan\theta)]|^2-
|\mbox{ep}[i(y+x\tan\theta)]|^2\right\} \, .
\label{eq:ho}
\end{eqnarray}
Eq. (\ref{eq:ho}) is a small quantity due to the property of the
exponential-like function $\mbox{ep}(iy)$. The H-regions are
symmetrical triangles. Obviously, $v_y$ vanishes when close to the
bottom side of the triangle. Along the symmetry line of the triangle,
$x\tan\theta=\mbox{(integer)}\times d$, which implies
$|\mbox{ep}[i(y\pm x\tan\theta)]|$ are equal; thus $v_y=0$. At
other points of these triangles, $v_y$ takes on small values, which
allow a smooth connection of the trajectories to the other regions.

So far we have shown clearly the formation of the bouncing
ball trajectories in the downstream transmitted region. Let's now
discuss the reflected region. From Eq. (\ref{eq:1}), (\ref{eq:C}) and
$C_{1n}=\delta_{nn_i}$, one has
 \begin{eqnarray}
\Psi_1&=&e^{ik_x x}\sin(k_n y)
+\sum_n(C_{2n}-C_{1n})e^{-i\alpha_n x}\sin(k_n y)
\nonumber\\
&=&2i\sin(k_x x)\sin(k_n y) + \Psi_2(-x,y)\nonumber\\
&\equiv& \Psi_R+\Psi_2(-x,y)
\end{eqnarray}
Here $\Psi_R$ is a standing wave solution for the total
reflection problem as if the channel were completely closed at
$x=0$; it represents a standing wave in the upstream or left part
of the channel. Thus the complete wave function upstream is a
superposition of this standing wave and a reflection of the
transmitted wave. The standing wave has a grid of nodal lines
which divides the region into cells of size $\Delta{x} =
\pi/k_x$, $\Delta{y} = \pi/k_y$. The magnitude of the standing
wave at the center of each cell is $\pm 2$. This pattern
will be preserved in $\Psi_1$ at the places not covered by
$\Psi_2(-x,y)$. The velocity in these places, however, will be
virtually zero since $\Psi_R$ is pure imaginary.

At the places covered by $\Psi_2(-x,y)$, the interference with
the standing wave can form quantum vortices\cite{Wu93}. These
quantum vortices perturb the stream lines, causing many
detours away from straight line trajectories. This is why the
trajectories upstream are much more complex than those downstream
in Fig.~1. Quantum vortices appear in many electron waveguide
structures, for examples, in that of Lent\cite{Lent}, and
Berggren\cite{Berggren}.

We now explain the deployment of the quantum vortices. To see a
quantum vortex, one must look for an isolated node of the
wavefunction, i.~e.  $\Psi_1/(2i)=\sin(k_x x)\sin(k_y y) +
\Psi_2(-x,y)/(2i)=0$. Thus $\Psi_2(-x,y)/(2i)$ must be real. Writing
$\Psi_2(-x,y)/(2i)$ as $R_2 \exp(iS/\hbar)$, we see that the nodes
must lie on curves of constant phase $S=\mbox{integer}\times h/2$. Now
${\bf v}=\nabla S/m^*$ would be the velocity if only $\Psi_2(- x,y)$
were present, so constant $S$ curves are orthogonal to ${\bf v}$.
Taking a region where the flow is to the right and downward, the
orbits are straight lines at an angle $-\theta$ from the
horizontal. Thus the constant $S$ curves are also straight lines but
at an angle $\pi/2-\theta$ upward. This is precisely the direction of
the upward diagonal line of the standing wave cell.  Picking a grid
point of the standing wave, $x=-(\pi/k_x)n_x$, $y=(\pi/k_y)n_y$, in
the region we are considering now,
\begin{eqnarray}
\Psi_2(-x,y)/(2i)&=&
-(1/4)\exp[i(k_x+k_y\tan\theta)\frac{\pi}{k_x}n_x]\times\nonumber\\
&&\mbox{ep}[i(\frac{\pi}{k_y}n_y-\frac{\pi}{k_x}n_x\tan\theta)]
\end{eqnarray}
Now remembering that $\mbox{ep}(iy)$ behaves as $\exp(ik_y y)$,
 \begin{equation}
\Psi_2(-x,y)/(2i)\approx -R_2 (-1)^{n_x+n_y}
\end{equation}
which is real on these grid points. Therefore, the constant phase
lines allowing quantum vortices are simply the upward diagonal
lines of the standing wave cells. The line passing through the
grid point $(n_x,n_y)$ is
 \begin{equation}
y-\frac{\pi}{k_y}n_y=\tan(\frac{\pi}{2}-\theta)(\frac{\pi}{k_x}n_x+x)
=\frac{k_x}{k_y} (\frac{\pi}{k_x}n_x+x) \, .
\end{equation}
Thus along the upward diagonal line,
 \begin{equation}
\sin(k_x x)\sin(k_y y)=(-1)^{n_x+n_y} \sin^2(k_x x) \, .
\end{equation}
Therefore the condition to find a node is
 \begin{equation}
R_2=\sin^2(k_x x) \, .
\end{equation}
Of course, the solution will depend on the form of $R_2$.  Once
again, since $\mbox{ep}(iy)\approx \mbox{constant}\times
\exp(ik_y y)$ in its non-vanishing region, $R_2$ is a slowly
varying function in the said region. Its average value along the
diagonal line can be estimated from the transmission coefficient
which yields $1/|4\cos\theta|$, a value normally less than 1.
Along the diagonal and within one cell, $\sin^2(k_x x)$ starts
from zero at one end, monotonically increases to 1, then
monotonically decreases to zero at the other end. Thus one
normally sees two vortices per cell along the diagonal direction
perpendicular to the average direction of motion of the particle.
Fig. 3 shows an array of such cells with two quantum vortices
spinning in the opposite directions. The deviations of the vortex
centers from the diagonal lines are small. Combined with Fig. 1,
one sees that the trajectories {\it on average} are mirror images of
those in the transmitted region which resemble classical
trajectories, but in detail they are much more complex.  We have
therefore found a complete justification for the ballistic
trajectory, or the ray optical picture of propagation in the
large wave number limit.  Corrections to this picture will
account for diffraction on both edges of the striped propagation
pattern, and slow filling in of the shadow regions on both sides
of the classical trajectory. This diffractive effect is seen in
Fig. 1, and it is much bigger when the energy is lowered, as seen
in Fig. 4. Notice that the average trajectories in the reflected
region close to the classical trajectory rely heavily on the fact
that the quantum vortices are situated according to the standing
wave cell structure. The effect of a small misalignment of
quantum vortices may be magnified significantly in the
trajectories. This is because there is an equal number of
hyperbolic unstable fixed points for each pair of
vortices, and these can cause two orbits which are
infinitesimally close in one place to follow quite different
paths later on. This effect is also seen in Fig. 4.

We would like to add the following additional points of discussion:

\begin{itemize}
\item
A natural question is, why do we not see turbulence despite the
very complex flow pattern? The answer is that we are solving a
steady flow problem. By definition, the velocity field does not
vary with time. Since the stream lines do not cross each other, a
single line can occupy very limited space. However, it would be
interesting to see what would happen if the steady state condition
were lifted.

\item
We have seen that as the energy is lowered, the trajectories
start to deviate from the classical ones. An extreme is reached
when the energy is so low that there is only one propagating
mode. In this case, the wavefunction is well described by the
single mode approximation\cite{Wu92} such that the transverse
dependence is simply $\sin(\pi y/d)$ except in the region very
close to the narrow constriction. Without the phase changing in
the $y$-direction, the stream lines are parallel to the wire. As
the stream lines approach the constriction, they will converge so
that they may pass through the opening. After escaping, they
then relax back and resume their motion parallel to the wire.
This situation is common to all devices with straight leads, at
low energy.

\item Since the stream lines do not cross each other, the bouncing
ball trajectories do not really hit the walls. Rather, they turn
to a more or less horizontal direction in the H-region( see Fig.
2), to avoid crossing. This is one difference from the strict
classical prediction.

\item A larger opening width $a$ will enlarge the H-region, and
thus lead to stream lines more parallel to the wire as expected.

\item Previously we said that on the up-stream side, where not
covered by $\Psi_2(-x,y)$, there is no flow motion. This is an
approximation. In reality, localized flow circulating in these
regions can form quantum vortices, and possibly vortex clusters.
This is a typical situation where stream lines close on
themselves.

\item The problem solved in this paper can be taken as an odd
symmetry solution for a wire of width $2d$ with a obstacle placed
at its center, which is a variation of the ordinary double slits
problem.  Or as an odd-symmetry solution for a wire with a
constriction standing out from each side of the channel.
\end{itemize}

In summary, we have given a quantitative description of
trajectories in a particular quantum wire transmission problem.
The result closely resembles the classical bouncing ball orbits
in the large wave number limit. The major difference lies in
the effect of quantum vortices which is responsible for the
complex pattern in the up-stream. This simple model can
serve as an explicit analytical example of the picture of
ballistic transport in quantum waveguides, which in one or
another form, clearly or vaguely, exists in our minds.

\acknowledgements

We are grateful to NSERC Canada for continued support under
research grant OGP00-3198.

\begin{figure}
\caption{Stream lines participating in global transmission for a
quantum wire with a simple constriction. $a=0.4d$, $E=50.5^2
(\pi/d)^2$, $n_i=40$. The angle $\theta$ reduced is
$52.4^\circ$.}
\end{figure}

\begin{figure}
\caption{Formation of the bouncing ball pattern in the transmitted
region due to the shift of the function $\mbox{ep}(\pm iy)$ as $x$
increases.}
\end{figure}

\begin{figure}
\caption{Current density plot in a portion of the up-stream
propagating region. The grid standing wave pattern and associated
quantum vortices are clearly seen. Horizontal and vertical lines
are standing wave node lines. Inclined diagonal lines are the
theoretical constant phase lines where $\Psi(-x,y)/(2i)$ is real.
Parameters are the same as Fig. 1. }
\end{figure}

\begin{figure}
\caption{Same as Fig.~1 but at a lower energy: $E=10.5^2 (\pi/d)^2$,
$n_i=8$.
}
\end{figure}

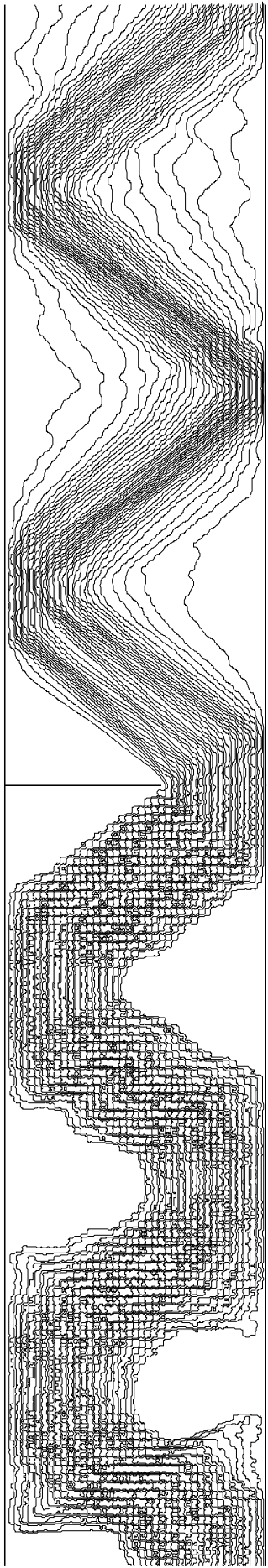
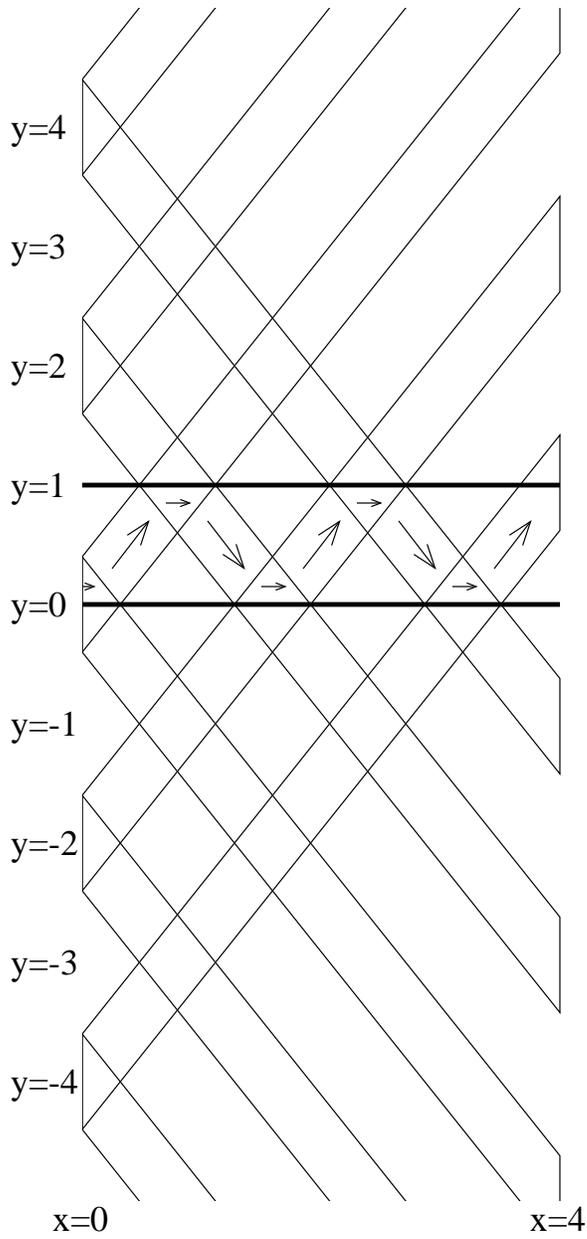
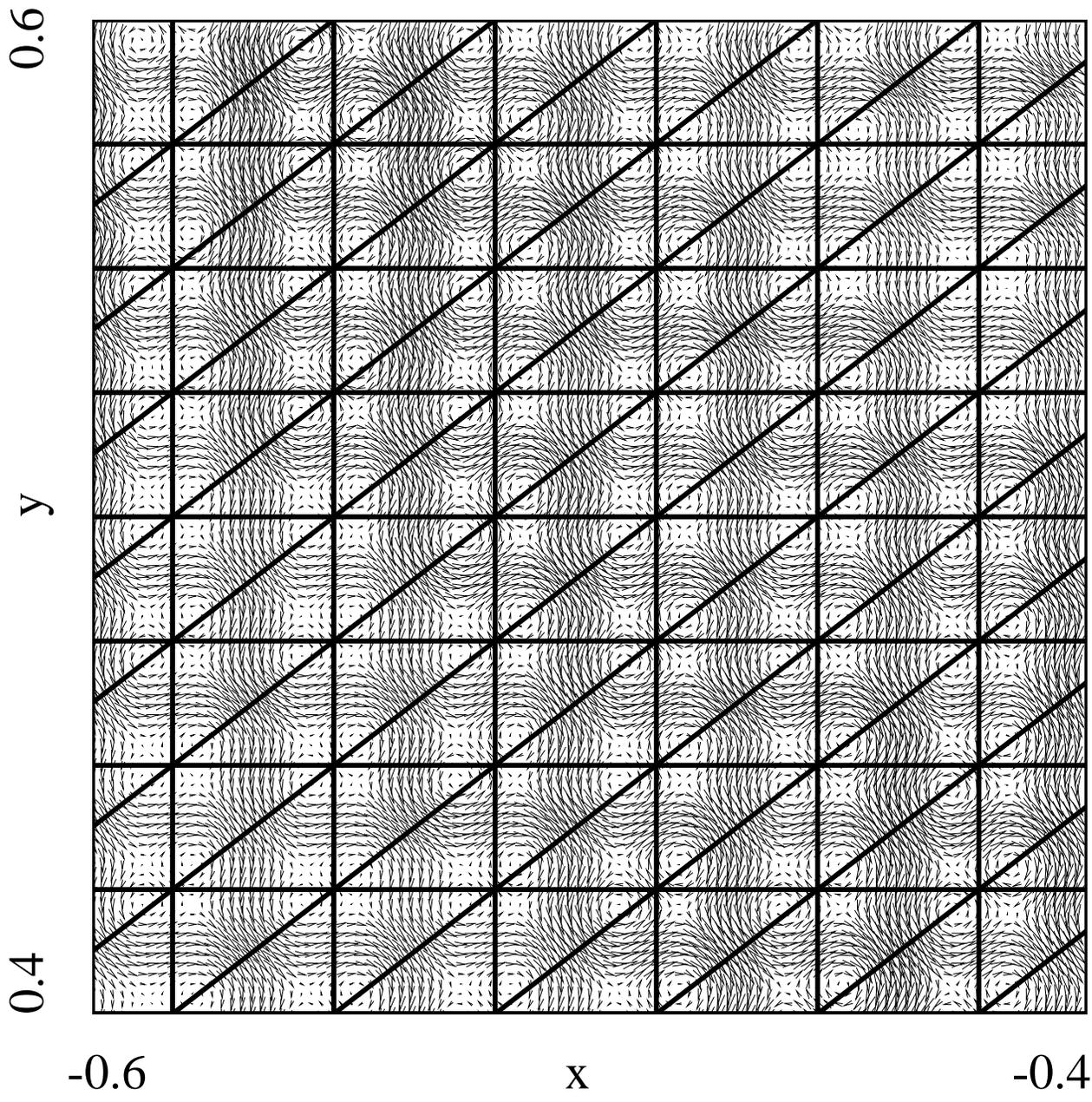
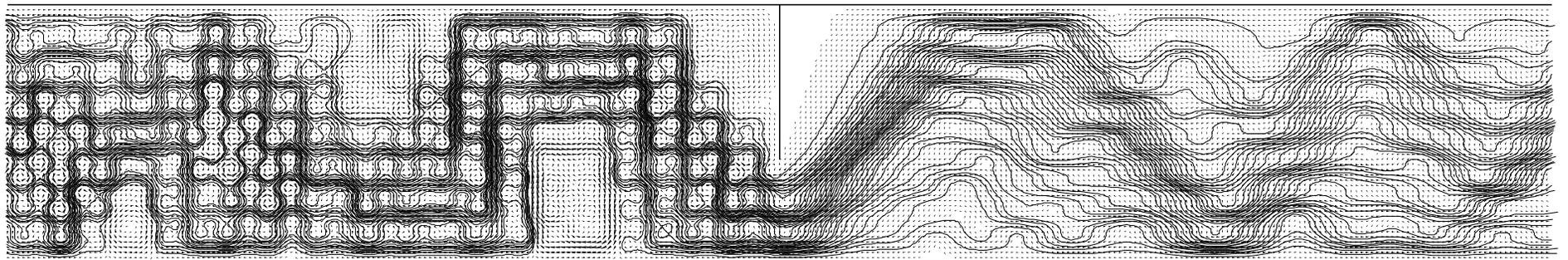
\end{document}